\documentclass[a4paper,USenglish]{lipics-v2019}
\usepackage[utf8]{inputenc}
\usepackage{nccmath}
\usepackage{fancyvrb}
\usepackage{xcolor}

\definecolor{symbolcolor}{rgb}{0.7, 0.1, 0.1}   
\definecolor{commentcolor}{rgb}{0.4, 0.4, 0.4}   
\definecolor{keywordcolor}{rgb}{0.0, 0.0, 0.8}    
\definecolor{varcolor}{rgb}{0.7, 0.2, 0.7}      

\newcommand{\an}{/\textbackslash}
\newcommand{\bd}[1]{\textcolor{symbolcolor}{#1}}
\newcommand{\vr}[1]{\textcolor{varcolor}{#1}}
\newcommand{\kw}[1]{\textcolor{keywordcolor}{#1}}

\title{Specifying verified x86 software from scratch}

\titlerunning{Specifying verified x86 software}

\author{Mario Carneiro}{Carnegie Mellon University, Pittsburgh, PA, USA}{mcarneir@andrew.cmu.edu}{https://orcid.org/0000-0002-0470-5249}{}

\authorrunning{M. Carneiro}

\Copyright{Mario Carneiro}


\category{}

\supplement{The formalization is a part of the Metamath Zero project at \url{https://github.com/digama0/mm0}.}

\funding{This material is based upon work supported by AFOSR grant FA9550-18-1-0120 and a grant from the Sloan Foundation.}

\acknowledgements{I would like to thank my advisor Jeremy Avigad for his support and encouragement, and for his reviews of early drafts of this work.}

\nolinenumbers 

\hideLIPIcs  

\EventEditors{John Harrison, John O'Leary, and Andrew Tolmach}
\EventNoEds{3}
\EventLongTitle{10th International Conference on Interactive Theorem Proving (ITP 2019)}
\EventShortTitle{ITP 2019}
\EventAcronym{ITP}
\EventYear{2019}
\EventDate{September 9--12, 2019}
\EventLocation{Portland, OR, USA}
\EventLogo{}
\SeriesVolume{141}
\ArticleNo{19}

\begin{document}

\maketitle

\begin{abstract}
We present a simple framework for specifying and proving facts about the input/output behavior of ELF binary files on the x86-64 architecture. A strong emphasis has been placed on simplicity at all levels: the specification says only what it needs to about the target executable, the specification is performed inside a simple logic (equivalent to first-order Peano Arithmetic), and the verification language and proof checker are custom-designed to have only what is necessary to perform efficient general purpose verification. This forms a part of the Metamath Zero project, to build a minimal verifier that is capable of verifying its own binary. In this paper, we will present the specification of the dynamic semantics of x86 machine code, together with enough information about Linux system calls to perform simple IO.
\end{abstract}

\section{Introduction}
The field of software verification is on the verge of a breakthrough. With the DeepSpec program \cite{pierce2016}, and associated projects like the CompCert C compiler \cite{leroy2012} and the seL4 microkernel \cite{klein2009}, we are finally seeing the feasibility of large-scale end-to-end software verification. But if we are to build such complicated systems, one critical component is the theorem prover itself. To what degree can we place our trust in these components?

The obvious solution is to verify the theorem prover. The CakeML compiler \cite{kumar2014} has come the farthest toward this goal, with the verification of a complete ML compiler in HOL4, which is capable of compiling itself. As HOL4 is written in ML, this forms a successful bootstrap of the complete proof.

There is a fundamental problem with bootstrapping a theorem prover, though: if the theorem prover has a bug, the bug may be exploited to prove it does not have a bug. If theorem prover $A$ is used to prove theorem prover $B$ correct, then $A$ must be trusted, even if $A=B$. Certainly a bootstrap provides a much greater correctness guarantee, but it remains a part of the trusted base even then. The best we can do is to decrease the size of the bootstrap until it can be easily inspected.

The Metamath Zero (MM0) project is an attempt to build a minimal full stack general-purpose verifier capable of verifying itself. By ``full stack'' we mean that the program is verified down to the lowest level with formal semantics, and ``minimal'' is measured with respect to the entire trusted part of the bootstrap: the verifier binary, and the statement of the correctness theorem. The key observation is that neither a compiler nor the proof of correctness need be trusted. Assuming the trusted part is correct, we know that the proof is checked correctly and the program that is output is described in the specification, so we know that the output program is also a correct verifier.

For this project, we targeted the x86-64 architecture on Linux. This is certainly not minimal in an absolute sense, but it is the lowest we can go on widely available hardware, which is important for replication.

While the verified MM0 verifier is still under construction, there is a Haskell reference implementation\footnote{\url{https://github.com/digama0/mm0/tree/master/mm0-hs}} that has been used to check the files in this report. MM0 is a logical framework, meaning that the axiom system is defined as part of the specification. Our goal is to formalize claims about a binary executable against the axioms of Peano Arithmetic, so the first part of the specification file\footnote{\url{https://github.com/digama0/mm0/blob/master/examples/peano.mm0}} defines PA together with operations on lists and bitvectors and sets such that we can usefully talk about instruction set semantics, running to 311 lines. The more difficult part is to specify the semantics of the binary itself, which we now turn to.

\section{The x86 Specification}

Our formalization of the x86 spec itself\footnote{\url{https://github.com/digama0/mm0/blob/master/examples/x86.mm0}} is based on the Sail x86 specification \cite{armstrong2018}. It is not a complete specification, but we do not need the complete semantics of x86 in order to specify our simple program. (This is a major problem with the compositional approach; because one layer does not know how the layer above will use it, it cannot skip anything. So the overall ``full-stack'' proof produced by putting everything together may be much larger than it needs to be.)

We also considered the K framework x86 specification \cite{dasgupta2019}, but struggled with the size of the specification. The Sail formalization was 1600 lines, but the K specification was distributed across several thousand files, including one file for each addressing mode / opcode combination, with a large and redundant flags specification generated automatically through fuzzing techniques. Particularly because the size of the specification is a part of the ``trusted part'' of our bootstrap, it was important that every part of the spec justify its presence and all redundancy is compressed away, to keep everything human readable. So the Sail formalization turned out to be just what we needed, and the fact that it was incomplete and hand-written were advantages, not disadvantages, in this context.

Because the MM0 verifier bootstrap is not yet complete, we also translated the Sail formalization into Lean,\footnote{\url{https://github.com/digama0/mm0/blob/master/mm0-lean/x86.lean}} so we have three formalizations of essentially the same specification in Sail, Lean, and MM0, and can do a bit of comparative analysis. Sail is an ML-like language with some weak dependent types for handling bitvectors, designed for specifying instruction set architectures. Lean is a full dependent type theory with support for inductive types and pattern matching. By comparison to these MM0 is laughably weak: it is based on multi-sorted first order logic (with no higher order types), and it has no syntax for inductive types or pattern matching. Not counting general bit operations and similar library code, looking only at the x86-specific part of the formalization, the specifications are about 1600 lines in Sail, 766 lines in Lean, and 1095 lines in MM0. We are heartened by this, because it shows we can hit the same order of magnitude as comparable formalizations in these other languages without any built-ins at all.

The formalization covers the decoding of bytes in memory into a sequence of instructions, covering most common user-mode instructions on integer registers. The dynamic semantics are specified as a relation on configurations: $\mathsf{step}\subseteq\mathsf{Config}\times\mathsf{Config}$, where $\mathsf{Config}$ is the set of tuples $(RIP, regs, flags, mem)$ such that $RIP\in u64$ is a 64-bit register (the instruction pointer), $regs:u4\to u64$ is a mapping giving the values of the 16 general purpose integer registers, $flags\in u64$ is another 64-bit register (only 4 bits of which we track), and $mem:u64\to (\mathsf{Prot}\times u8)$ defines the values of the virtual memory of the application. Here $\mathsf{Prot}=u3$ are the read/write/execute bits associated to pages of virtual memory; the semantics ensure that loads only read from readable bytes and writes are to writable bytes.

Here is an example statement from the MM0 specification, defining the execution behavior of the \texttt{mul} instruction:
\begin{Verbatim}[commandchars=\\\{\}]
\kw{theorem} execXASTMul \{\bd{n src res lo hi}: nat\} (\vr{k sz rm k2}: nat):
  $ execXAST \vr{k} (xastMul \vr{sz rm}) \vr{k2} <-> E. \bd{n} E. \bd{src} E. \bd{res} (
      \bd{n} = wsizeBits \vr{sz} \an
      readEA \vr{k} \vr{sz} (RM_EA \vr{k rm}) \bd{src} \an
      \bd{res} = readRegSz \vr{k} \vr{sz} RAX * \bd{src} \an
      ifp (\bd{n} = 8)
        (\vr{k2} = writeReg \vr{k} wSz16 RAX \bd{res})
        (E. \bd{lo} E. \bd{hi} (
          splitBits ((\bd{n} <> \bd{lo}) : (\bd{n} <> \bd{hi}) : 0) \bd{res} \an
          \vr{k2} = writeReg (writeReg \vr{k sz} RAX \bd{lo}) \vr{sz} RDX \bd{hi}))) $;
\end{Verbatim}
The statement enclosed in dollar delimiters is a theorem statement. Although we cannot describe the syntax in detail here, we note the traditional operators of first order logic: `\texttt{<->}' is `if and only if', `\texttt{E.}' is `exists', and `\texttt{\an}' is `and'. The free variables in the statement are purple and the bound variables are red; notice that all variables have type \texttt{nat}, because PA is untyped---all variables range over natural numbers. The `\texttt{<>}' operator is the pairing function, `\texttt{:}' is the cons operator, and `\texttt{*}' is the usual multiplication of natural numbers in PA.

In the Sail and Lean formalizations, this is one branch of a large pattern match. Since MM0 does not have pattern matching, we need to find a way to write essentially the same definition without a lot of boilerplate, especially in places where we could make an unchecked error. The trick we use to avoid needing to pack the entire definition into one statement is through a definition specification:

\begin{Verbatim}[commandchars=\\\{\}]
\kw{def} execXAST (\vr{k ast k2}: nat): wff;
\end{Verbatim}
This command declares the existence of a definition, but doesn't say what it is. Later theorems, like \texttt{execXASTMul}, refer to the definition, and a proof of the specification must provide a definition that satisfies all subsequent theorems. (This is similar to a Coq \textsf{Module}.)

And what of types? Sail and Lean both use strong type systems, which have long been used for simple error checking in programming languages, and by using an untyped theory we are apparently giving this up. But we can recover types by putting them explicitly into the language as predicates:
\begin{Verbatim}[commandchars=\\\{\}]
\kw{theorem} execXASTT (\vr{k ast k2}: nat):
  $ \vr{k} e. Config \an execXAST \vr{k ast k2} -> \vr{ast} e. XAST \an \vr{k2} e. Config $;
\end{Verbatim}
This also allows us to capture input/output variables in multiple argument predicates such as this.

\section{IO and the ELF specification}
The \texttt{execXAST} function from the last section is the main part of the \texttt{step} relation, which loads an instruction from RIP and executes it. This relation is nondeterministic where the Intel manual leaves parts unspecified, or when we wish to abstract from the detailed behavior that is specified (particularly if it depends on a part of the state outside our user-mode execution model). Conversely, when a state contains behavior that we wish to avoid (such as triggering a segmentation or protection fault), or if it invokes an instruction that we have not specified, then the step relation will not step to anything from that state.

The one exception to this is in IO commands. We support only the \texttt{syscall} instruction for making system calls to Linux, but in order to model the results we extend the state. We let $\textsf{KernelState}=\textsf{List}\;u8\times \textsf{List}\;u8\times\textsf{Config}$, where the two lists represent input waiting on \texttt{stdin} and output written to \texttt{stdout}. Note that these are unbounded lists, which are not meant to be realistically representable in the hardware; since these are streams they don't exist in memory at all and are produced through user interaction. Nevertheless, they allow us to turn an x86 program into an abstract function on byte streams.

Finally, we define the specification of the ELF format. As we are doing no linking we need only a single program header containing the code to execute and no section headers, simplifying the spec considerably. The initial memory is nondeterministic, except for a small stack allocation containing the command line arguments, and with the code loaded in memory. The program is given access to the \texttt{mmap()} system call for requesting memory, and \texttt{read()} and \texttt{write()} for IO. Altogether, this requires an additional 237 lines of MM0 code.

\section{Conclusion}
With this work we have demonstrated that it is possible to build nontrivial specifications on an extremely spartan framework. As long as the framework is sufficiently extensible, we can have everything we want at the same time: a tiny and efficient trusted verifier, verifying everything about our programs down to the metal, in a weak logic, with clearly delimited theorem statements that are only as complicated as the endpoint specification itself.

\bibliography{references}

\begin{thebibliography}{1}

\bibitem{armstrong2018}
Alasdair Armstrong et~al.
\newblock {Detailed Models of Instruction Set Architectures: From Pseudocode to
  Formal Semantics}.
\newblock In {\em Proceedings of the 25th Automated Reasoning Workshop},
  page~13, 2018.

\bibitem{dasgupta2019}
Sandeep Dasgupta, Daejun Park, Theodoros Kasampalis, Vikram~S Adve, and Grigore
  Ro{\c{s}}u.
\newblock A complete formal semantics of x86-64 user-level instruction set
  architecture.
\newblock In {\em Proceedings of the 40th ACM SIGPLAN Conference on Programming
  Language Design and Implementation}, pages 1133--1148. ACM, 2019.

\bibitem{klein2009}
Gerwin Klein et~al.
\newblock {seL4: Formal verification of an OS kernel}.
\newblock In {\em Proceedings of the ACM SIGOPS 22nd symposium on Operating
  systems principles}, pages 207--220. ACM, 2009.

\bibitem{kumar2014}
Ramana Kumar, Magnus~O Myreen, Michael Norrish, and Scott Owens.
\newblock {CakeML: a verified implementation of ML}.
\newblock In {\em ACM SIGPLAN Notices}, volume~49, pages 179--191. ACM, 2014.

\bibitem{leroy2012}
Xavier Leroy et~al.
\newblock {The CompCert verified compiler}.
\newblock {\em Documentation and user’s manual. INRIA Paris-Rocquencourt},
  53, 2012.

\bibitem{pierce2016}
Benjamin~C. Pierce.
\newblock {The Science of Deep Specification (Keynote)}.
\newblock In {\em Companion Proceedings of the 2016 ACM SIGPLAN International
  Conference on Systems, Programming, Languages and Applications: Software for
  Humanity}, SPLASH Companion 2016, pages 1--1, New York, NY, USA, 2016. ACM.
\newblock \href {http://dx.doi.org/10.1145/2984043.2998388}
  {\path{doi:10.1145/2984043.2998388}}.

\end{thebibliography}

\end{document}